\documentclass[cameraready]{Interspeech}

\title{AV-EMO-REASONING: BENCHMARKING EMOTIONAL REASONING CAPABILITIES IN
OMNI-MODAL LLMS WITH AUDIO-VISUAL CUES}

\author[affiliation={2}]{Dingkun}{Zhou}
\author[affiliation={1}]{Krish}{Patel}
\author[affiliation={1}]{Ajay}{Kankipati}
\author[affiliation={1}]{Akshaj}{Gupta}
\author[affiliation={1}]{Zeyi Austin}{Li}
\author[affiliation={1}]{Mohul}{Shukla}
\author[affiliation={1}]{Vibhor}{Narang}
\author[affiliation={1}]{Sara}{Kofman}
\author[affiliation={3}]{Zongli}{Ye}
\author[affiliation={1}]{Grace}{Wang}
\author[affiliation={1}]{Xiaoyu}{Shi}
\author[affiliation={1}]{Tingle}{Li}
\author[affiliation={4}]{Guan-Ting}{Lin}
\author[affiliation={1}]{Kan Jen}{Cheng}
\author[affiliation={5}]{Huang-Cheng}{Chou}
\author[affiliation={1}]{Jiachen}{Lian}
\author[affiliation={1}]{Gopala}{Anumanchipalli}

\address{
$^1$ UC Berkeley, USA \
$^2$ South China University of Technology, China \
$^3$ Zhejiang University, China \
$^4$ National Taiwan University, Taiwan \
$^5$ University of Southern California, USA
}

\email{}
\keywords{speech recognition, human-computer interaction, computational paralinguistics}

\usepackage{comment}
\usepackage[percent]{overpic} 
\newcommand{\mypar}[1]{\noindent{\bf #1}}
\begin{document}

\maketitle

\begingroup
\renewcommand\thefootnote{\fnsymbol{footnote}}
\footnotetext[1]{Dingkun Zhou and Krish Patel contributed equally. Jiachen Lian is the project lead and corresponding author: \texttt{jiachenlian@berkeley.edu}.}
\setcounter{footnote}{0}
\endgroup

\begin{abstract}
    Emotions conveyed through voice and face shape engagement and context in human–AI interaction. Despite rapid progress in omni-modal large language models (LLMs), the holistic evaluation of emotional reasoning with audiovisual cues remains limited. To address this gap, we introduce AV-EMO-Reasoning, a benchmark designed to systematically assess emotional reasoning abilities in LLMs. The framework leverages a curated audiovisual corpus comprising synthetic single- and multi-turn dialogues and a real-world subset, together with emotion-perception and interaction-reasoning metrics, to assess whether LLMs can appropriately understand users’ emotions and produce appropriate responses. By releasing a systematic evaluation benchmark, AV-EMO-Reasoning offers a reproducible standard for evaluating emotion-aware dialogue and advances toward more natural, adaptive human–AI interaction.
\end{abstract}

\section{Introduction}
\label{sec:introduction}
Human communication encompasses more than the words we say.
Speech meaning is fundamentally shaped by vocal tone, emotional prosody, and temporal dynamics \cite{quinto2013emotional, mullennix2002effects}.
When auditory cues are flat or ambiguous, non-verbal signals such as facial expressions, gaze, and head motion often provide the necessary context to decode a speaker's intent \cite{busso2004analysis}.
While Large Language Models (LLMs) \cite{brown2020language} and Spoken Dialogue Models (SDMs) \cite{fang2024llama, xie2024mini, wang2024freeze, zeng2024glm, defossez2024moshi} have made significant strides in parsing linguistic content and basic acoustic affect, they often overlook the fluid, cross-modal nature of real world interaction.
Emotional states are not static; they shift dynamically throughout a conversation.
For instance, a smile may accompany apologies, and facial expressions can convey critical information during silence.
Capturing, tracking, and responding to these subtle shifts across both voice and face is essential for building modern systems that are intuitive, natural, and empathetic.

While many SDMs process linguistic content well, they frequently overlook the rational use of emotion \cite{liu2025emo}, such as acknowledging distress with empathy or amplifying joy to build rapport. 
Prior efforts have introduced benchmarks for spoken question answering \cite{hassid2023textually, shih2023gsqa, lin2024align}, instruction following \cite{huang2024dynamic, huang2024dynamic2}, style-conditioned conversation \cite{ao2024sd, lin2024advancing, lin2024can, lin2024paralinguistics}, duplex systems \cite{lin2025full, lin2025full2, liu2025emo}, facial expression \cite{schuhmann2025emonet, zhao2023dfme}, talking head generation \cite{chen2020comprises, zhou2024subjective}, and recognizing affect~\cite{lian2023av} from speech or video alone. These works address important facets of dialogue, yet they do not systematically evaluate whether an SDM or, more broadly, an omni-modal LLM can detect, interpret, and incorporate a user’s emotional state using both audio and visual cues in a context-aware manner.

\begin{figure}[t]
\centering
\begin{overpic}[width=0.48\textwidth]{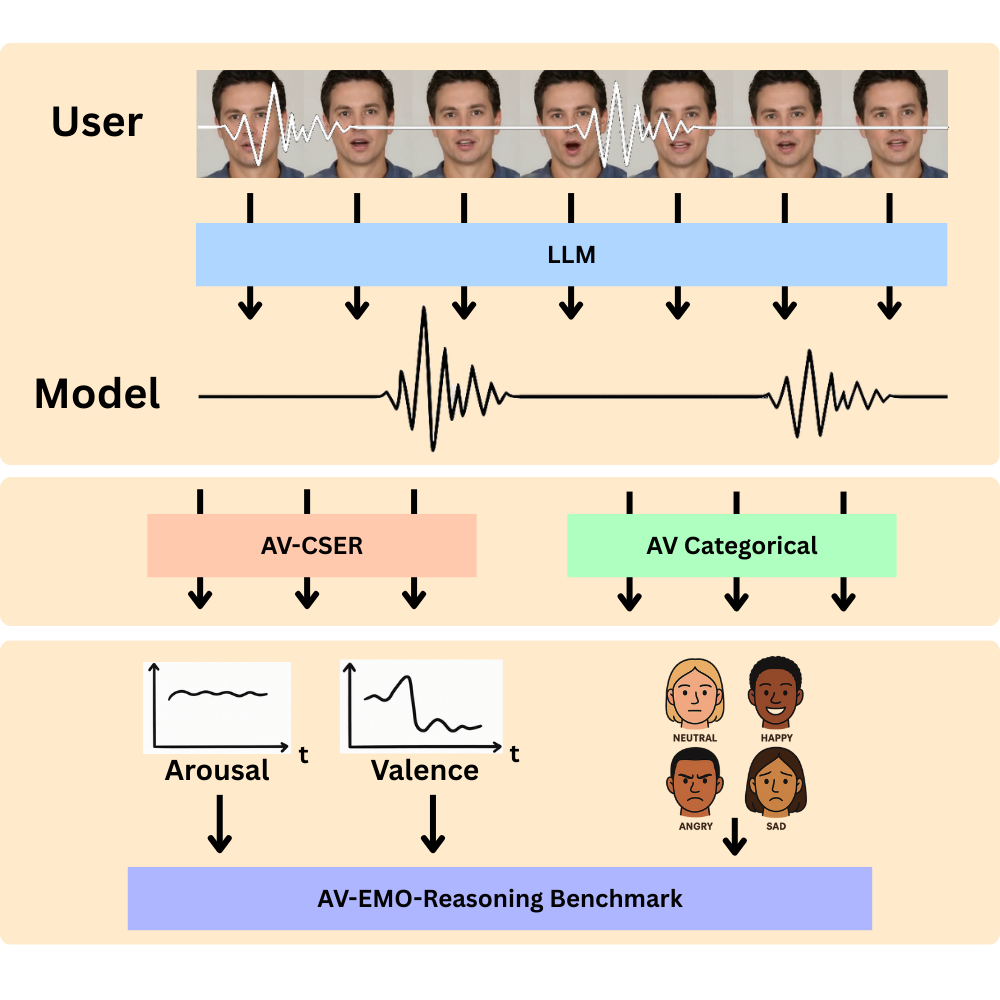}
  \put(98,98){\footnotesize\bfseries} 
\end{overpic}
\vspace{-5mm}
\caption{The AV-EMO-Reasoning Framework.}
\label{fig:avemo}

\end{figure}


Speech-only evaluation is insufficient for emotion reasoning since many affective cues are ambiguous in voice, complementary across modalities, or visible yet silent \cite{Goncalves_2025}. 
The nuance of real-world conversation often hinges on the congruence between these expressive streams.
For instance, sarcasm, masking, or specific cultural speaking styles can flatten prosody, leaving facial action patterns as the primary reliable indicators of intent.
Furthermore, since vocal arousal and facial valence frequently diverge, perceiving them jointly significantly improves inference accuracy, particularly in environments characterized by acoustic noise or channel mismatch.
Some emotional markers such as tears, micro-expressions, and gaze aversion often leave little to no acoustic trace, yet they remain visually self-evident.
Because humans are highly sensitive to cross-modal incongruence \cite{10890558} such as a smile paired with a cold, detached tone and adjust their social expectations accordingly \cite{wang2024cross}, computational frameworks must account for these multi-modal discrepancies to achieve affective alignment.
Restricting the evaluation to speech leaves these complex, high-stakes communicative phenomena untested.

To bridge this gap, we propose AV-EMO-Reasoning (Figure~\ref{fig:avemo}), a benchmark that evaluates an omni-modal LLM’s ability to detect, interpret, and integrate a user’s emotional state from vocal tone, linguistic content, and visual signals, and effectively reflect this understanding in its responses. We evaluate both discrete emotion categories (e.g., happy, angry) and continuous affective trajectories to capture a broad range of affective phenomena across turns.

Concretely, AV-EMO-Reasoning evaluates two complementary dimensions: emotion understanding and emotion reasoning. Emotion understanding measures whether LLMs can correctly recognize a speaker’s emotion under multimodal inputs, remain robust when emotional modalities are missing, and resolve audio-visual conflicts. Emotion reasoning measures whether LLMs can track, regulate, and respond to emotional dynamics across turns. These metrics also reveal the internal patterns and preferences in how they interpret emotions.
To facilitate fine-grained temporal analysis, we developed an Audio-Visual Continuous Emotion Recognition (AV-CSER) model. Unlike standard turn-level classifiers, AV-CSER generates frame-level estimates for arousal and valence that are precisely aligned with both raw waveforms and video frames. This continuous stream complements our discrete turn-level labels, providing a more granular ground truth for evaluation.
To enable fine-grained continuous analysis, we develop an audio-visual continuous emotion recognition (AV-CSER) model that produces frame-level arousal/valence estimates aligned to both waveform and video frames, complementing a discrete classifier for turn-level labels. Our experiments indicate that current omni-modal LLMs underperform on audio-visual emotion reasoning. Their performance degrades sharply in the absence of explicit acoustic cues, and in multi-turn dialogue they tend to mirror the user’s emotional state rather than regulating or guiding them appropriately. 
These findings underscore a substantial opportunity for future works to improve through more integrated, joint reasoning across modalities. 
We summarize our contributions as follows:


\mypar{Audio-visual dialogue data.} We collect an emotionally expressive audio-visual dialogue set with single-turn and multi-turn interactions, aligning speech with facial expressions to enable synchronized evaluation.

\mypar{Audio-visual continuous emotion analysis.} We train an AV-CSER model for frame-level arousal/valence trajectories from audio and video, supporting detailed temporal studies of affect.

\mypar{Cross-modal understanding and reasoning metrics.} We introduce two categories of metrics, understanding and reasoning, to test whether the model can correctly interpret emotions and reliably use cross-modal cues to guide how the dialogue unfolds.

\mypar{Benchmarking omni-modal LLMs.} We provide AV-EMO-Reasoning, a new benchmark for emotional reasoning in omni-modal LLMs/SDMs, and report baseline results that surface key failure modes and directions for improvement.

\section{AV-EMO-Reasoning Framework}
\label{sec:method}
 We evaluate LLMs along two dimensions: emotion understanding and emotion reasoning. All metrics are normalized to the range $[0,1]$. We use four high-agreement emotion categories (neutral, happy, angry, and sad) to keep discrete evaluation reliable and balanced. Finer-grained emotional variation is captured by continuous arousal and valence measures and by human perceptual ratings.

\subsection{Emotion Understanding Metrics}
We evaluate our metrics on both single-turn and multi-turn dialogue segments. For multi-turn segments, each segment contains 3–5 consecutive dialogue turns, and we fix the first speaker in the segment as the user. Given the full dialogue context, we then independently evaluate the emotion of each user turn in the segment; in the single-turn setting, we directly evaluate the emotion for that turn. For each user turn, we measure the model’s emotion understanding ability under four input conditions: Full-AV, A-only, V-only, and Conflict. For each user turn, we measure the model's emotion understanding ability under four input conditions: Full-AV for categorical emotion understanding, A-only and V-only for missing-modality robustness, and Conflict for audio-visual conflict preference.

\subsubsection{Categorical Emotion Understanding (Full-AV)}
We first evaluate the model's basic ability to recognize the user's emotion at each dialogue turn under the full audio-visual condition. \textbf{$\text{Acc}_\text{AV}$} measures the proportion of user turns whose predicted emotion matches the ground-truth label. Concretely, we count how many user turns have matching predictions and annotations under the Full-AV input and divide this by the total number of user turns. Full-AV macro-averaged $\textbf{F1}_\text{AV}$ evaluates whether the model's performance is balanced across emotion categories. We compute the F1 score for each category separately and then average them. The emotion label set includes four categories: neutral, happy, angry, and sad.

\subsubsection{Missing Modality Robustness (A/V-Only)}
To quantify the robustness of emotion understanding under missing-modality conditions, we evaluate the model's emotion classification accuracy with \textbf{A-only} and \textbf{V-only} inputs. In the A-only setting, the visual modality is replaced with repeated copies of the first frame of the original video. We count how many user turns have predicted emotion labels that match the human-annotated ground-truth labels under the A-only condition and divide this number by the total number of user turns. Similarly, in the V-only setting, the speech signal is replaced with noise. We count how many user turns have predicted emotion labels that match the human-annotated ground-truth labels under the V-only condition and divide this number by the total number of user turns.

\subsubsection{AV-Conflict Preference}
In user turns where the audio and video convey conflicting emotions, we further analyze which modality the model tends to rely on under cross-modal conflict.
\textbf{Audio-Align Rate (AAR)} measures the proportion of conflict turns for which the model’s predicted emotion matches the audio emotion label. Concretely, we count how many conflict turns have predictions equal to the audio label and divide this number by the total number of conflict turns.
\textbf{Visual-Align Rate (VAR)} and \textbf{Other-Label Rate (OLR)} are defined analogously for the visual label and for predictions that match neither label.

\subsection{Emotion Reasoning Metrics}
Our emotion Reasoning metrics are adapted from the EMO-Reasoning\cite{liu2025emo} and are evaluated uniformly on both single-turn and multi-turn dialogue segments. All continuous metrics are defined in the arousal--valence (V/A) space, and we group them into single-turn and multi-turn variants. A single-turn metric applies a function to a local V/A trajectory. A multi-turn metric aggregates these single-turn scores across dialogue turns and reflects how the model maintains, regulates, and guides emotion over the full conversation.

\subsubsection{Single-turn Metrics}
ECS measures how promptly the model perceives and follows the user’s emotional changes. EBS measures how strongly the model’s trajectory converges toward a balanced trajectory when the user is in an extreme state. ESS measures the smoothness of the emotional trajectory within a turn, penalizing sharp fluctuations. ERS is the average of ECS, EBS, and ESS and reflects overall single-turn emotion understanding and regulation.

\subsubsection{Multi-turn Metrics}
CT-ECS averages ECS over all turns in a multi-turn dialogue. It evaluates the model's overall emotional contagion ability across the full conversation.
CT-EBS averages EBS only over turns where the user shows an extreme emotional state. It evaluates the model's global ability to balance emotion throughout the dialogue.
CT-ESS extends ESS to the multi-turn setting. It first applies dynamic time warping between the V/A trajectories of adjacent turns and then normalizes and aggregates the distances. It evaluates the continuity and smoothness of emotional trajectories across turns.
CT-ERS is computed by averaging CT-ECS, CT-EBS, and CT-ESS. It evaluates the model's overall multi-turn emotion reasoning ability.
\subsection{Subjective Human Perception Evaluation}
We conducted a human evaluation using ratings from annotators who successfully passed a pre-training session. Annotators assessed 15 dialogues from each of the three LLMs and the baseline, rating each response for rationality, naturalness, and relevance on a 5-point Likert scale. Each dialogue was rated by at least five annotators. The collected Likert scores were then averaged and normalized to the range [0,1], where higher values indicate better quality.
The details of the metrics are below.

\mypar{Emotional Rationality (ER)} judges how appropriate the model’s emotion is to the user’s state. Not just “matching,” it rewards understanding and empathy.

\mypar{Emotional Naturalness (EN)} rates how natural and smooth the emotional expression of the SDM sounds, free of robotic, forced, or mechanical tone.

\mypar{Response Relevance (RR)} scores how well the content addresses the user’s emotional and contextual needs; detailed, specific, context-aware replies rate higher than generic platitudes.

\section{Categorical A-V-Emotion Recognition}
\label{sec:categorical}
\mypar{Model and Dataset.}
Our audio-visual emotion detection model is trained with random video modality dropout, yielding a unified framework capable of inference on both multi-modal (audio-video) and uni-modal (audio-only) data.
Our model is trained on the MultiDialog dataset \cite{multidialog}, a corpus of face-to-face, audio-visual dialogues with per-utterance emotion annotations. The dataset comprises 8,733 dialogues (340 hours) from 12 speakers. We adhere to the official split, using 7,011 dialogues for training, 891 for validation, and 831 for testing.

\mypar{Model Configuration and Baseline.}    
To establish single-modality baselines, we extracted audio embeddings using SenseVoice-Small \cite{sensevoice} and facial embeddings using EmotiEffLib \cite{savchenko2023facial}. As shown in Table~\ref{tab:Weighted_F1_scores}, our proposed multi-modal approach significantly outperforms both the audio-only and video-only baselines on the MultiDialog dataset \cite{multidialog}.

    \begin{table}[!b]
    \vspace{-3mm}
    \fontsize{7}{9}\selectfont
    \centering
      \caption{\small This table summarizes the Weighted F1 scores for all models evaluated on the MultiDialog dataset \cite{multidialog}.}
      \vspace{-2mm}
      \label{tab:Weighted_F1_scores}
      \begin{tabular}{@{}lllll@{}}
        \toprule
        Model & Neutral & Happy & Angry & Sad \\ \midrule
        SenseVoice Small \cite{sensevoice} & 0.4745 & 0.4599 & 0.3717 & 0.3128 \\
        EmotiEffLib \cite{savchenko2023facial} & 0.5642 & 0.3033 & 0.0639 & 0.0727 \\
        \textbf{Ours} & \textbf{0.6074} & \textbf{0.6350} & \textbf{0.6074} & \textbf{0.5111} \\ \bottomrule
      \end{tabular}
    \end{table}



\begin{figure*}[h]
  \centering
  \setlength{\tabcolsep}{4pt}
  \begin{tabular}{@{}ccc@{}}
    \includegraphics[width=0.25\textwidth]{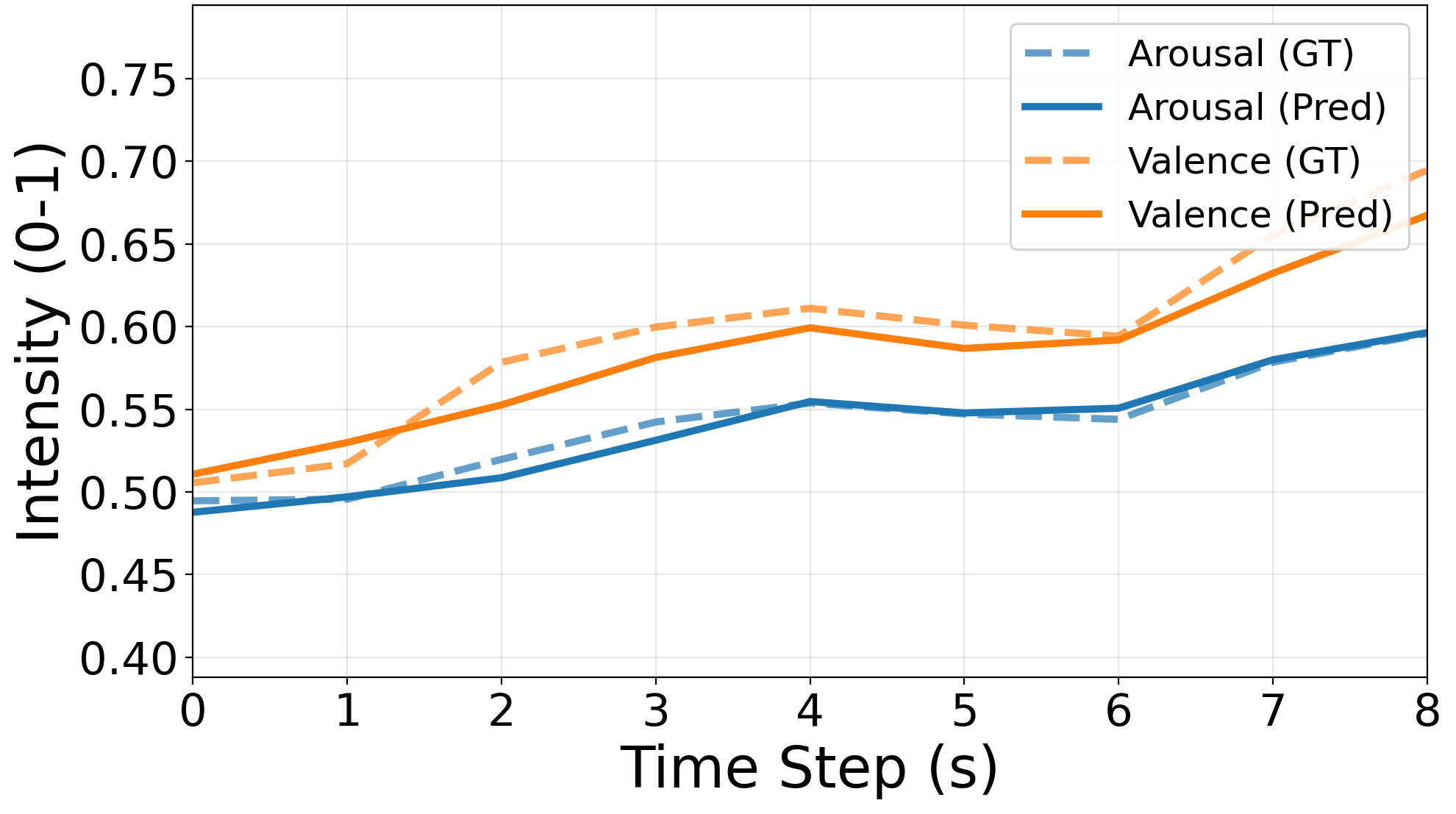} &
    \includegraphics[width=0.25\textwidth]{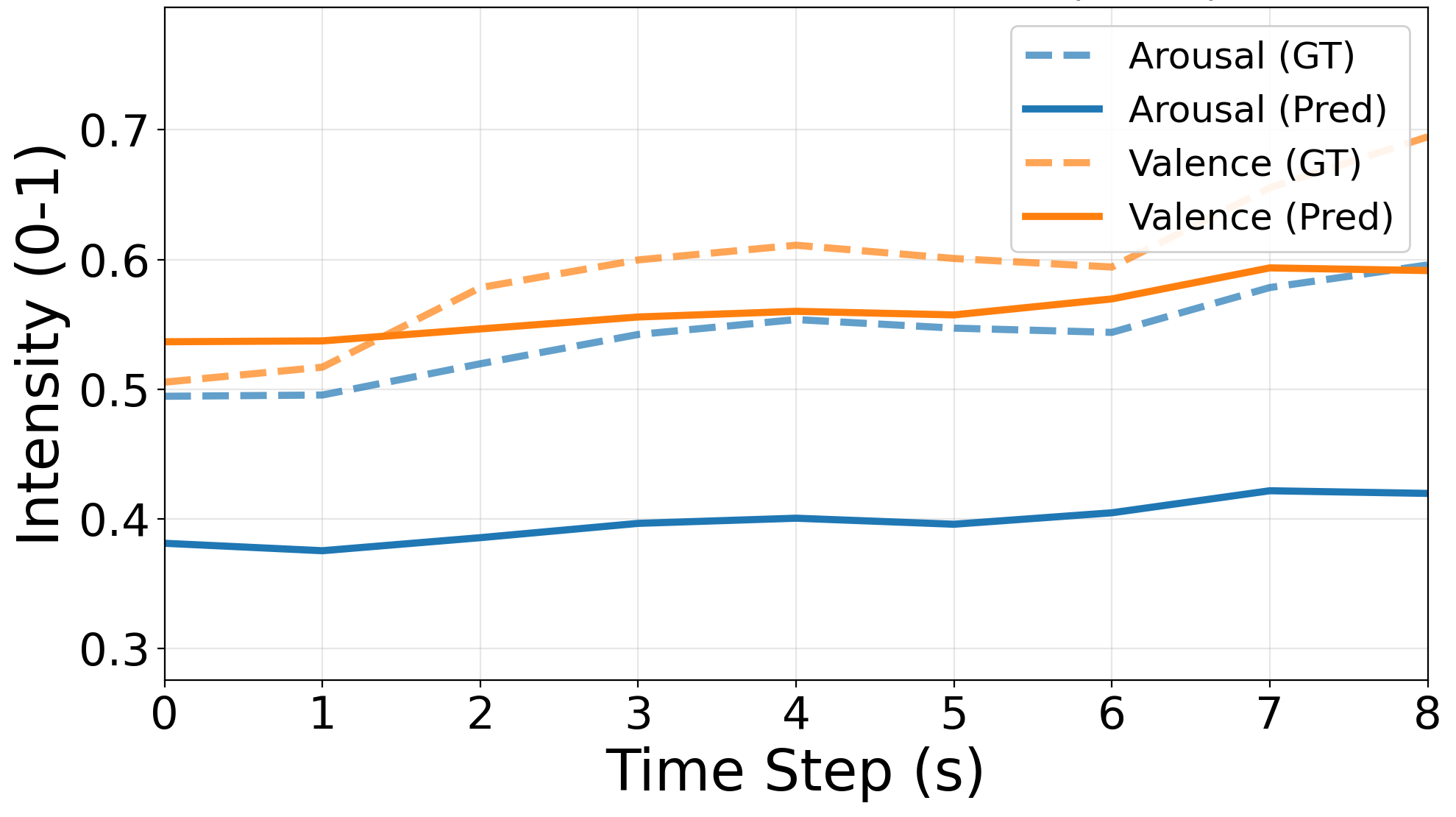} &
    \includegraphics[width=0.25\textwidth]{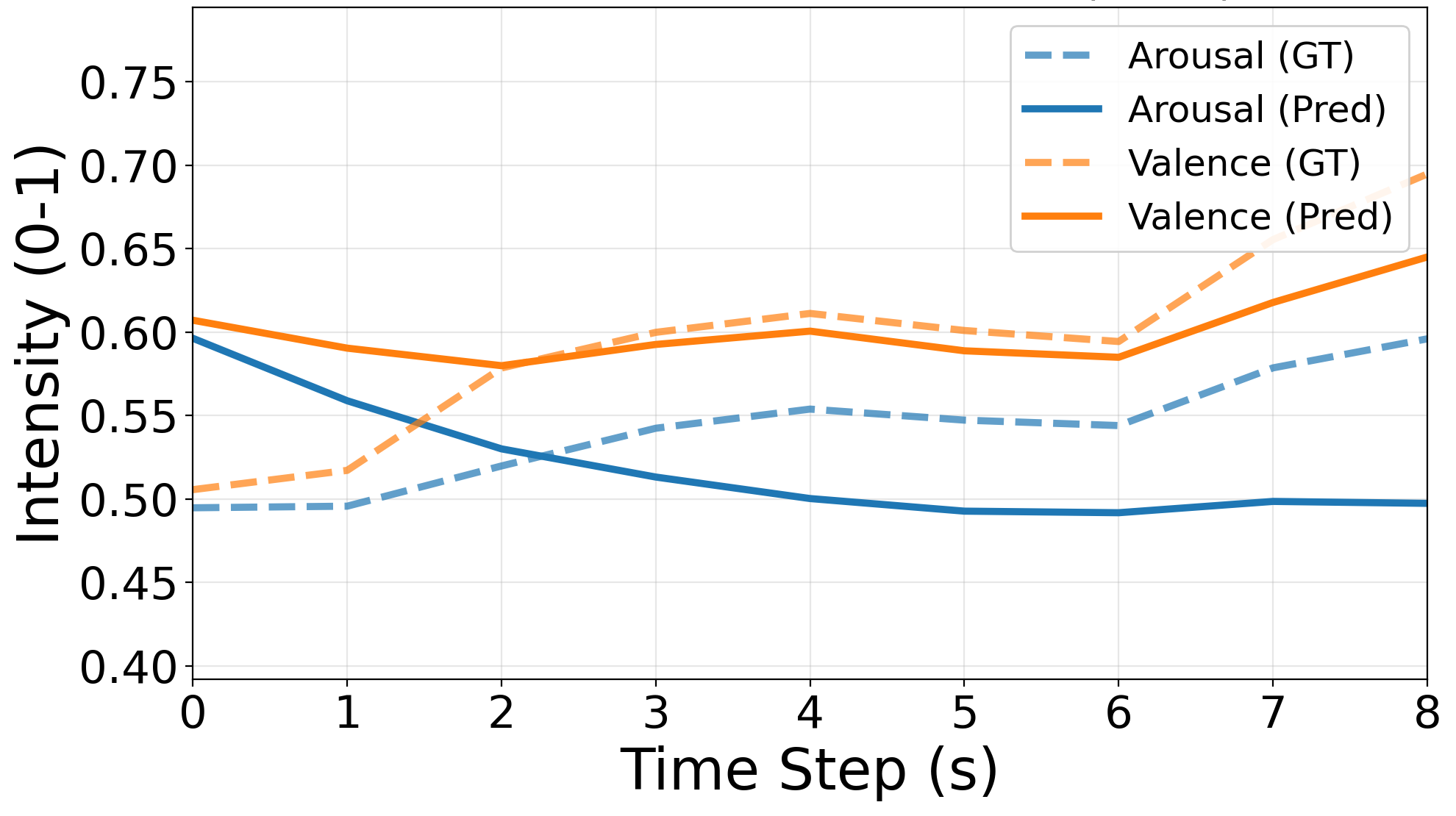} \\
    \footnotesize Audio + Video & \footnotesize Audio & \footnotesize Video
  \end{tabular}
  \vspace{-3mm}
  \caption{\small Continuous emotion predictions across modalities on the RECOLA.}
  \label{fig:recola_triptych}
  \vspace{-2mm}
\end{figure*}

\begin{table*}[!t]
\scriptsize
\centering
\caption{\small Results on the Synthetic (single-/multi-turn) and the MultiDialog (multi-turn) datasets.}
\label{tab:all_results_stacked}
\vspace{-2mm}
\centering
\setlength{\tabcolsep}{3pt}
\begin{tabular}{@{}l
  *{4}{S[table-format=1.4]}
  *{7}{S[table-format=1.3]}
  *{4}{S[table-format=1.3]}
@{}}
\toprule
& \multicolumn{4}{c}{\textbf{Emotion Reasoning}} & \multicolumn{7}{c}{\textbf{Emotion Understanding}} & \multicolumn{4}{c}{\textbf{Human Evaluation}} \\
\cmidrule(lr){2-5}\cmidrule(lr){6-12}\cmidrule(l){13-16}

{\textbf{Model (Single-Turn, Synthetic)}} &
\multicolumn{1}{c}{ECS} & \multicolumn{1}{c}{EBS} &
\multicolumn{1}{c}{ESS} & \multicolumn{1}{c}{ERS} &
\multicolumn{1}{c}{$\text{Acc}_\text{AV}$} &
\multicolumn{1}{c}{$\text{F1}_\text{AV}$} &
\multicolumn{1}{c}{$\text{Acc}_\text{A}$} &
\multicolumn{1}{c}{$\text{Acc}_\text{V}$} &
\multicolumn{1}{c}{$\text{AAR}$} &
\multicolumn{1}{c}{$\text{VAR}$} &
\multicolumn{1}{c}{$\text{OLR}$} &
\multicolumn{1}{c}{ER} & \multicolumn{1}{c}{EN} &
\multicolumn{1}{c}{RR} & \multicolumn{1}{c}{ERS} \\
\midrule
Synthetic (GPT-4 + CosyVoice) &
0.767 & 0.672 & 1.000 & 0.737 &
\multicolumn{7}{c}{—} &
0.590 & 0.535 & 0.573 & 0.567 \\
\midrule
Baichuan 1.5-o \cite{li2025baichuan} &
0.577 & 0.483 & 1.000 & 0.524 &
\bfseries 0.841 & 0.311 & 0.547 & 0.760 & 0.490 & \bfseries 0.195 & 0.270 &
0.628 & 0.575 & 0.618 & 0.607 \\
MiniCPM-o 2.6 \cite{yao2024minicpm} &
0.660 & 0.492 & 1.000 & 0.589 &
0.804 & \bfseries 0.343 & \bfseries 0.865 & 0.824 & \bfseries 0.686 & 0.104 & 0.264 &
\bfseries 0.888 & \bfseries 0.873 & \bfseries 0.950 & \bfseries 0.903 \\
Qwen 2.5-o \cite{Qwen2.5-Omni} &
\bfseries 0.731 & \bfseries 0.587 & 1.000 & \bfseries 0.672 &
0.713 & 0.243 & 0.627 & \bfseries 0.850 & 0.632 & 0.137 & \bfseries 0.317 &
0.665 & 0.720 & 0.710 & 0.698 \\

\addlinespace[2pt]\midrule

{\textbf{Model (Multi-Turn, Synthetic)}} &
\multicolumn{1}{c}{CT-ECS} & \multicolumn{1}{c}{CT-EBS} &
\multicolumn{1}{c}{CT-ESS} & \multicolumn{1}{c}{CT-ERS} &
\multicolumn{1}{c}{$\text{Acc}_\text{AV}$} &
\multicolumn{1}{c}{$\text{F1}_\text{AV}$} &
\multicolumn{1}{c}{$\text{Acc}_\text{A}$} &
\multicolumn{1}{c}{$\text{Acc}_\text{V}$} &
\multicolumn{1}{c}{$\text{AAR}$} &
\multicolumn{1}{c}{$\text{VAR}$} &
\multicolumn{1}{c}{$\text{OLR}$} &
\multicolumn{1}{c}{ER} & \multicolumn{1}{c}{EN} &
\multicolumn{1}{c}{RR} & \multicolumn{1}{c}{ERS} \\
\midrule
Synthetic (GPT-4 + CosyVoice) &
0.747 & 0.385 & 0.838 & 0.623 &
\multicolumn{7}{c}{—} &
0.860 & 0.830 & 0.785 & 0.825 \\
\midrule
Baichuan 1.5-o \cite{li2025baichuan} &
0.619 & 0.039 & 0.806 & 0.426 &
0.763 & 0.307 & 0.598 & 0.806 & 0.694 & 0.028 & \bfseries 0.321 &
0.768 & 0.695 & 0.748 & 0.737 \\
MiniCPM-o 2.6 \cite{yao2024minicpm} &
0.667 & 0.161 & 0.657 & 0.445 &
0.728 & 0.347 & \bfseries 0.673 & 0.797 & \bfseries 0.796 & 0.021 & 0.213 &
\bfseries 0.850 & \bfseries 0.793 & \bfseries 0.868 & \bfseries 0.837 \\
Qwen 2.5-o \cite{Qwen2.5-Omni} &
\bfseries 0.730 & \bfseries 0.162 & \bfseries 0.814 & \bfseries 0.526 &
\bfseries 0.815 & \bfseries 0.384 & 0.520 & \bfseries 0.821 & 0.600 & \bfseries 0.114 & 0.266 &
0.768 & 0.740 & 0.790 & 0.766 \\

\addlinespace[2pt]\midrule

{\textbf{Model (Multi-Turn, Real)}} &
\multicolumn{1}{c}{CT-ECS} & \multicolumn{1}{c}{CT-EBS} &
\multicolumn{1}{c}{CT-ESS} & \multicolumn{1}{c}{CT-ERS} &
\multicolumn{1}{c}{$\text{Acc}_\text{AV}$} &
\multicolumn{1}{c}{$\text{F1}_\text{AV}$} &
\multicolumn{1}{c}{$\text{Acc}_\text{A}$} &
\multicolumn{1}{c}{$\text{Acc}_\text{V}$} &
\multicolumn{1}{c}{$\text{AAR}$} &
\multicolumn{1}{c}{$\text{VAR}$} &
\multicolumn{1}{c}{$\text{OLR}$} &
\multicolumn{1}{c}{ER} & \multicolumn{1}{c}{EN} &
\multicolumn{1}{c}{RR} & \multicolumn{1}{c}{ERS} \\
\midrule
Real (MultiDialog \cite{multidialog}) &
0.749 & \multicolumn{1}{c}{0.623} & 0.788 & 0.394 &
\multicolumn{7}{c}{—} &
0.540 & 0.538 & 0.480 & 0.519 \\
\midrule
Baichuan 1.5-o \cite{li2025baichuan} &
0.632 & \multicolumn{1}{c}{0.407} & \bfseries 0.765 & \bfseries 0.362 &
0.720 & 0.287 & 0.580 & 0.770 & 0.670 & 0.030 & \bfseries 0.300 &
0.378 & 0.508 & 0.355 & 0.413 \\
MiniCPM-o 2.6 \cite{yao2024minicpm} &
0.648 & \multicolumn{1}{c}{\textbf{0.501}} & 0.688 & 0.258 &
0.760 & \bfseries 0.363 & \bfseries 0.720 & 0.760 & \bfseries 0.760 & 0.020 & 0.220 &
\bfseries 0.580 & \bfseries 0.568 & 0.558 & 0.568 \\
Qwen 2.5-o \cite{Qwen2.5-Omni} &
\bfseries 0.741 & \multicolumn{1}{c}{0.406} & 0.747 & 0.341 &
\bfseries 0.770 & \bfseries 0.363 & 0.550 & \bfseries 0.790 & 0.640 & \bfseries 0.110 & 0.250 &
0.578 & 0.555 & \bfseries 0.685 & \bfseries 0.601 \\
\bottomrule
\end{tabular}
\vspace{-5mm}
\end{table*}

\section{Continuous A-V-Emotion Recognition}

\label{sec:continuous}

To evaluate our proposed metrics, we train an audio-visual continuous speech emotion recognition (AV-CSER) model that extracts fine-grained temporal emotion trajectories from synchronized speech and video signals. We use RECOLA dataset~\cite{ringeval2013introducing} for training. For each segment, we use HiCMAE~\cite{sun2024hicmae} to pre-extract features and downsample them to two values per second, namely valence and arousal. The regression head is a single-layer BiLSTM with a hidden size of 256, and its outputs are mapped to the range $[0,1]$ using a sigmoid function. During training, we optimize the concordance correlation coefficient (CCC) loss with a learning rate of $1\times10^{-3}$, and apply a \texttt{ReduceLROnPlateau} scheduler.

Our results show that multimodal fusion consistently outperforms unimodal methods in all settings. Detailed comparisons with the prior audio-only CSER baseline are reported in Table~\ref{tab:cser_model_comparisons}, while modality-specific results across subject-wise and official RECOLA splits are reported in Table~\ref{tab:recola_ccc_two_splits}. These performance trends are also visualized in Figure~\ref{fig:recola_triptych}, which compares the model predictions with the ground-truth trajectories (dashed lines) for a sample segment. The fused A+V model most accurately tracks changes in both arousal and valence, capturing sharp emotional shifts around $t \approx 2$--4\,s and $t \approx 7$--8\,s with almost no delay. In contrast, the audio-only model can follow short-term changes in arousal but shows a systematic negative bias of about 0.08--0.12 and noticeably attenuated amplitudes. The video-only model aligns well with the valence level but underfits the arousal trajectory and exhibits a larger initial offset at the beginning of the sequence.
\begin{table}[!t]
\scriptsize
\centering
  \caption{\small Concordance Correlation Coefficient (CCC) comparison between our method and previous audio-only approach for RECOLA (subject-wise split).}
  \label{tab:cser_model_comparisons}
  \vspace{-2mm}
  \label{tab:CCC}
  \centering
    \begin{tabular}{@{}lllll@{}}
        \toprule
        \textbf{Model} & Arousal & Valence & Mean \\ \midrule
        CSER \cite{liu2025emo} & 0.403 & 0.164 & 0.284 \\
        \textbf{Ours}    & \textbf{0.687} & \textbf{0.690} & \textbf{0.689} \\ 
        \bottomrule
    \end{tabular}
    \vspace{-3mm}
\end{table}

\begin{table}[!b]
\vspace{-3mm}
\fontsize{7}{9}\selectfont
\centering
\caption{\small Concordance Correlation Coefficient (CCC) across modalities on the RECOLA for subjective-wise and official splits.}
\vspace{-2mm}
\label{tab:recola_ccc_two_splits}
\setlength{\tabcolsep}{4pt}
\begin{tabular}{@{}l *{3}{S[table-format=1.3]} *{3}{S[table-format=1.3]}@{}}
\toprule
& \multicolumn{3}{c}{\bf Subject-wise split} & \multicolumn{3}{c}{\bf Official split} \\
\cmidrule(lr){2-4} \cmidrule(l){5-7}
\textbf{Modality} &
\multicolumn{1}{c}{Arousal} &
\multicolumn{1}{c}{Valence} &
\multicolumn{1}{c}{Mean} &
\multicolumn{1}{c}{Arousal} &
\multicolumn{1}{c}{Valence} &
\multicolumn{1}{c}{Mean} \\
\midrule
Audio & 0.282 & 0.139 & 0.211 & 0.191 & 0.039 & 0.115 \\
Video & 0.397 & 0.468 & 0.432 & 0.228 & 0.323 & 0.276 \\
\textbf{Audio + Video} & \bfseries 0.687 & \bfseries 0.690 & \bfseries 0.688 & \bfseries 0.644 & \bfseries 0.463 & \bfseries 0.553 \\
\bottomrule
\end{tabular}
\end{table}



\vspace{-3mm}

\section{Experimental Setup}
\vspace{-3mm}
\label{sec:experiment}


\subsection{Dataset}
\vspace{-1mm}
\mypar{Synthetic Data.}
We constructed our synthetic dataset using a three-stage pipeline. 
First, we prompted GPT-4 \cite{achiam2023gpt} to generate both single-turn and multi-turn dialogues. 
Each utterance was accompanied by a stylistic prompt describing the desired vocal delivery
Second, these text-and-prompt pairs were used as input for the CosyVoice text-to-speech model \cite{du2024cosyvoice} to synthesize high-quality emotional speech. 
Finally, for each audio segment, a random facial image was selected from the AI-Face dataset \cite{lin2025aiface} and animated using DreamTalk \cite{ma2023dreamtalk}. 

On top of this base, we further construct missing-modality and modality-conflict conditions. For the audio-missing condition, we replace the speech track with noise while keeping the original video unchanged. For the visual-missing condition, we replace the video with a simple temporal repetition of the first frame. To obtain strongly polarized modality-conflict samples, we synthesize a pair of audio-visual clips with opposite emotional polarity from the same transcript. Given an emotion label and its opposite, we construct two corresponding style prompts and generate the two clips. We then combine the audio from one clip with the video from the other so that the audio and visual streams carry explicitly opposite discrete emotion labels, forming AV-Conflict samples.

\mypar{Real Data.}
To evaluate model performance on real human speech, we use the MultiDialog dataset~\cite{multidialog}. The construction of the audio-missing (A-missing) and visual-missing (V-missing) conditions follows the same procedure as in the synthetic data setting. For the modality-conflict condition, we select pairs of video clips from the real dataset that share the same speaker ID, have opposite emotion labels, and differ in duration by no more than 1\,s. We then combine the audio from one clip with the video from the other, thereby constructing AV-Conflict samples in which the audio and visual streams carry explicitly opposite discrete emotion labels.

\mypar{LLM Response Collection.}
We use the same prompt to collect responses from three omni-modal language models, each supporting audio-visual inputs and native speech synthesis: Baichuan-Omni-1.5 \cite{li2025baichuan}, MiniCPM-o 2.6 \cite{yao2024minicpm}, and Qwen 2.5-Omni-7B \cite{Qwen2.5-Omni}. 
\vspace{-2mm}

\section{Results}
Table~\ref{tab:all_results_stacked} summarizes the benchmark results across emotion understanding, emotion reasoning, and human evaluation.

\mypar{Emotion Understanding Performance.}
The model can reliably identify common emotions but struggles to distinguish rare ones, and its performance drops substantially under audio-only input. Multimodal fusion does not clearly surpass the best single modality and is often dominated by a single channel. When emotional cues in audio and video conflict, AAR is typically much higher than VAR, indicating that the model tends to align with the audio label even though audio is not the more reliable modality overall. This preference for the weaker modality under conflict reveals a mismatch between the model's implicit modality weighting and the actual effectiveness of each modality on the task.

\mypar{Emotion Reasoning Performance.}
All models achieve ECS scores close to the reference trajectory, showing that they adjust their emotional trajectories in the same direction as the user, with Qwen exhibiting the strongest emotional contagion. In contrast, EBS is consistently below the baseline, suggesting that when the user is in an extreme emotional state, the models rarely pull their emotion back toward a more balanced range. In multi-turn dialogue settings, all models reach relatively high ESS scores and their trajectories are generally continuous and stable, although MiniCPM exhibits the largest fluctuations. Overall, in real multi-turn conversations, none of the current models has yet reached human-level emotional reasoning or regulation.

\mypar{Human Evaluation.}
On automatic metrics, the ground-truth real and synthetic dialogues show stronger affect dynamics and label consistency than LLM-generated responses. In contrast, on perceptual metrics, annotators prefer the LLM outputs, likely because they are more verbose and descriptive. Automatic scores further indicate that canonical dataset answers better preserve emotional tone across turns, even when humans favor the content of LLM-generated replies. This leaves substantial headroom for LLMs to sound truly humanlike: wording and phrasing are well received, but prosody, pacing, and affect still lag genuine human speech.
\section{Conclusion and Future Work}
\vspace{-1mm}
\label{sec:conclusion}
This paper introduces AV-Emo-Reasoning, a framework for evaluating LLMs' understanding and reasoning abilities in multimodal settings. We further build a joint audio-visual categorical emotion recognizer and an AV-CSER for fine-grained temporal analysis.Our experiments show that current omni-modal LLMs still exhibit a clear gap between recognizing emotion and speaking with emotionally appropriate responses. It remains necessary to better model the temporal dynamics of emotion and to design systems that can maintain and adapt their emotional state over the course of a full dialogue. We hope that AV-Emo-Reasoning will serve as a foundation for future LLM research on emotional perception and emotional reasoning.
\clearpage
\bibliographystyle{IEEEtran}
\bibliography{mybib}

\end{document}